\documentclass[prl, twocolumn, floatfix]{revtex4}
\setcitestyle{super}

\usepackage{graphicx}
\usepackage{hyperref}
\usepackage{amsmath}
\usepackage{color}

\begin{document}

\title{Instability threshold in a large balanced magneto-optical trap}

\author{M. Gaudesius, R.\ Kaiser and G.\ Labeyrie\footnote{To whom correspondence should be addressed.}}
\affiliation{Universit\'{e} C\^{o}te d'Azur, CNRS, Institut de Physique de Nice, 06560 Valbonne, France}
\author{Y. Zhang and T. Pohl}
\affiliation{Department of Physics and Astronomy, Aarhus University, DK 8000 Aarhus, Denmark}

\begin{abstract}

Large clouds of cold atoms prepared in a magneto-optical trap can develop spatio-temporal instabilities when the frequency of the trapping lasers is brought close to the atomic resonance. This system bears close similarities with trapped plasmas, whereby effective Coulomb interactions are induced by the exchange of scattered photons and lead to collective nonlinear dynamics of the trapped atoms. We report in this paper a detailed experimental study of the instability threshold, and comparisons with three-dimensional simulations of the interacting, laser-driven cloud.
\end{abstract}

\maketitle
\section{I. Introduction}
The magneto-optical trap (MOT), first demonstrated in 1987~\cite{Pritchard1987}, is widely used nowadays. It has triggered a broad field of research in the past decades. Following its initial demonstration, early work explored the role of sub-Doppler mechanisms~\cite{Cohen1989} and multiple scattering of light~\cite{Walker1990}. In addition to representing essential experimental technology to reach quantum degeneracy in ultracold gases, MOTs can harbor in the large atom number limit a variety of interesting nonlinear phenomena that are not yet fully understood and bear intriguing ties to plasma~\cite{Mendonca2008,Rodrigues2016,Barre2019, Hansen1975,Evrard1979} and stellar physics~\cite{Labeyrie2006, Cox1980}. Descriptions borrowed from the fields of nonlinear dynamics~\cite{Wilkowski2000, DiStephano2004} and fluid physics~\cite{Mendonca2012} have been employed to investigate these phenomena. 

In a MOT with a large number of trapped atoms $N$, these are subjected to three forces. A trapping and cooling force is exerted by laser beams in the presence of a magnetic field gradient. This force depends on laser and magnetic field parameters, but not on $N$. Two additional, \textit{collective} forces appear when $N$ is large enough. First, the laser beams are attenuated inside the atom cloud due to photon scattering. This attenuation yields a compressive correction to the trapping force~\cite{Dalibard1988}. Second, the scattered photons can be re-scattered by other atoms, which gives rise to a Coulomb-like repulsive force~\cite{Walker1990}. It is the interplay between these three forces that can generate unstable dynamics in large MOTs.

During the last two decades, instabilities in MOTs have been studied in various configurations~\cite{Wilkowski2000, DiStephano2003, DiStephano2004, Romain2016, Labeyrie2006}. In Ref.~\cite{Labeyrie2006} we reported an unstable behavior for a large balanced MOT (see section II) and presented a preliminary study of its instability threshold. A simple unidimensional analytical model allowed us to provide a rough instability criterion. In the present work, we provide a detailed analysis of the instability threshold using an improved experimental scheme where the number of trapped atoms is controlled. Some of our observations deviate from the scaling predicted by the analytical model of Ref.~\cite{Labeyrie2006}. We thus developed a three-dimensional numerical simulation based on microscopic theoretical ingredients, whose results are in qualitative agreement with the experiment.

The article is organized as follows. In section II, we describe our experimental setup and measurement procedure. We then review previous theoretical models of MOT instabilities in section III.a, and describe our numerical approach in section III.b. In section IV, we present our experimental results and discuss the comparison to our numerical simulations. The implications of our findings and ensuing perspectives for future work are outlined in sections V and VI.

\section{II. Experimental Setup}
\label{experiment}

\begin{figure}
\begin{center}
\includegraphics[width=1.0\columnwidth]{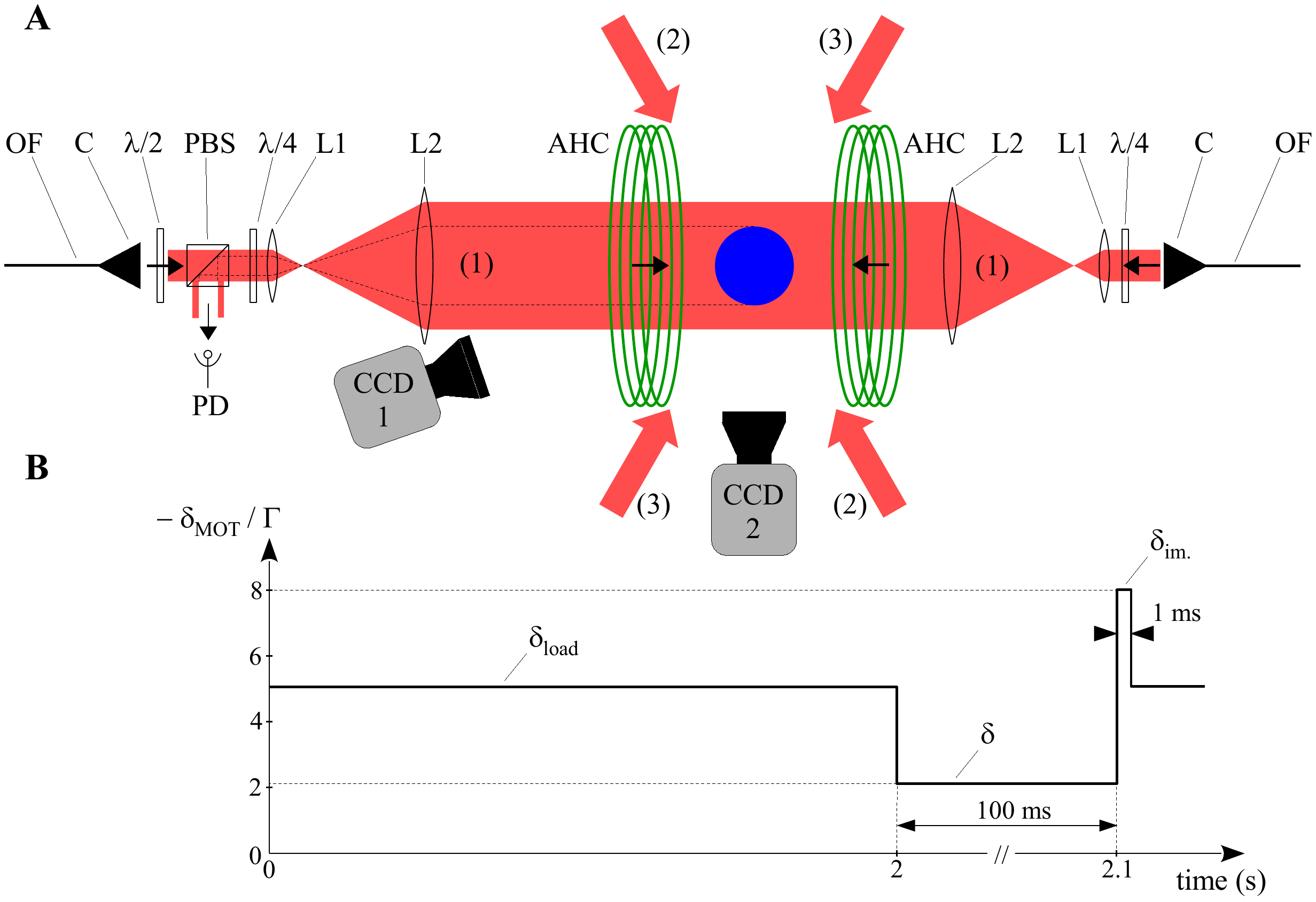}
\caption{Experimental procedure. \textbf{A} Details of the arrangement for one pair of MOT beams (1). The other two pairs ((2) and (3)) are identical and orthogonal to the first one. The beams are delivered by polarization-maintaining single-mode optical fibers (OF) coupled to a collimator (C). The beams intensities are balanced using a half-wave plate ($\lambda/2$) and polarizing beam splitter (PBS) assembly placed on one arm. The PBS also allows to collect part of the counter-propagating beam after its passage through the cloud, for optical density measurement using a photodiode (PD). The beams are expanded to a waist of 3.4 cm using afocal telescopes (L$_1$ + L$_2$). Their polarization is adjusted using quarter-wave plates ($\lambda/4$). The magnetic field gradient is provided by a pair of anti-Helmoltz coils (AHC). \textbf{B} Timing of the experiment. The MOT is loaded for 2 s with a detuning $\delta_{load}$, adjusted to maintain the number of atoms fixed during the measurement. The detuning is then changed to $\delta$ for 100 ms (instability phase). An image is finally acquired with a fixed detuning $\delta_{im.} = -8\Gamma$.}
\label{setup}
\end{center}
\end{figure}

Our large MOT and its characteristics have been thoroughly described in Ref.~\cite{Camara2014}. Here, we briefly reiterate the aspects that are most relevant to the present work. We use six large (3.4 cm waist) trapping laser beams of same intensity tuned close (detuning $\delta$) to the $F = 2 \rightarrow F^{\prime} = 3$ transition of the $^{87}$Rb D2 line to trap and cool atoms from an ambient vapor (see Fig.~\ref{setup}). These six beams, originating from the same source coupled into single-mode optical fibers, form three counter-propagating pairs crossing at 90$^\circ$. Because of this balanced arrangement, the nature of the instabilities is different from that of Refs.~\cite{Wilkowski2000,Romain2016} where the beams were retro-reflected, such that the center-of-mass motion played a dominant role in the nonlinear dynamics of the trapped atoms. In our setup, the intensities of the beams in each pair are balanced, yielding in principle a centro-symmetric situation. However, a source of asymmetry is the magnetic field gradient $\nabla$B, which is generated by a pair of anti-Helmoltz coils and is thus twice stronger along the coil's axis than in the transverse plane. Small defects in the spatial profiles of the beams are also creating a local intensity imbalance, which is another source of symmetry breaking. The peak intensity for each beam is typically $I = 5$ mW/cm$^2$. The corresponding saturation parameter per beam is $s = \frac{I/I_{sat}}{1+4(\delta/\Gamma)^2} \approx 0.08$ for $\delta = -3\Gamma$, assuming atoms are pumped into the streched Zeeman sub-states ($I_{sat} = 1.67$ mW/cm$^2$). $\Gamma$ is the natural line width of the transition ($\frac{\Gamma}{2 \pi} = 6.06$ MHz).

In addition to the trapping light, all beams also contain a small amount of ''repumping'' light tuned close to the $F = 1 \rightarrow F^{\prime} = 2$ transition to maintain the atoms in interaction with the trapping lasers. Three pairs of Helmoltz coils are used to compensate for stray magnetic fields at the position of MOT center. Because of this, the position of the cloud in the stable regime does not vary in the course of the experiment when the magnetic field gradient is adjusted.

In section IV, we will present the measured threshold detunings as the magnetic field gradient is varied, while keeping the number of trapped atoms $N$ fixed. Since we employ a vapor-loaded MOT where the steady-state value of $N$ is determined by ($\delta$, $\nabla$B), this requires to use the temporal sequence of Fig.~\ref{setup}\textbf{B}. It is composed of three successive phases that are continuously cycled. In the first phase, we load the MOT for a duration of 2 s with a detuning $\delta_{load}$. We set the value of $N$ by adjusting $\delta_{load}$. We then, in the second phase, rapidly change the detuning for 100 ms to a value $\delta$, which determines the dynamical regime of the MOT that we wish to probe. Finally, in the third phase, the detuning is adjusted to $\delta_{im.} = -8~\Gamma$ for 1 ms to perform a fluorescence image acquisition.

We image the cloud from two directions at 90$^\circ$, giving access to projections of the atomic distribution along the three spatial dimensions. Note that the assumption that the detected fluorescence is proportional to the column density is safe because of the large detuning chosen for the imaging~\cite{Camara2014}. An absorption imaging scheme, not represented in Fig.~\ref{setup}, allows us to determine the value of $N$.

The measurement sequence of Fig.~\ref{setup}B enables to maintain a fixed atom number determined by $\delta_{load}$. The 100 ms delay between detuning step and image acquisition is necessary to decorrelate the observed dynamics from the ''kick'' applied to the atoms as the detuning is abruptly changed. Note however that this delay is small compared to the loading time constant of the MOT ($\approx 2$ s), which ensures that the atom number is determined by $\delta_{load}$ and remains approximately independent of $\delta_{inst.}$. Since only one image is recorded for each sequence, by cycling over many runs (typically a hundred) we randomly probe the dynamics.

\section{III. Previous theoretical approaches and three-dimensional numerical simulations}

\subsection{III.a. Previous theoretical approaches}

We briefly recall here the evolution of the theoretical approaches that were employed in the past to describe the physics of large balanced MOTs. This evolution ultimately led to the development of the three-dimensional numerical approach described in Section III.b.

The model introduced in the 90s by C. Wieman and co-workers~\cite{Walker1990} was the first to describe the operation of the MOT in the stable, multiple-scattering regime. It relies on a Doppler description of the trapping forces which seems appropriate for very large MOTs, although it is known that sub-Doppler mechanisms can play an important role for alkali MOTs at low and moderate atom numbers~\cite{Cohen1989, Kim2004}.

Contrary to the case of small atom numbers, where the dynamics is governed by single-atom physics and the MOT size is determined by the temperature of the gas, in the regime of large atom numbers the radiation pressure forces acting on the atoms depend on the atomic density distribution and can therefore lead to collective behavior. This occurs when the optical density of the cloud at the trapping laser frequency becomes non-negligible. On one hand, the trapping beams are then attenuated while propagating through the cloud, which produces a density dependent compression force~\cite{Dalibard1988}. On the other hand, the scattered photons can be re-scattered by other atoms, resulting in a Coulomb-like repulsion force~\cite{Walker1990}. Because the absorption cross-section for scattered photons $\sigma_R$ is different from and larger than that for laser photons $\sigma_L$, the repulsion force is larger than the compression force and the cloud expands when $N$ is increased~\cite{Walker1990, Steane1992}. At equilibrium, the atomic density inside the cloud is constant and only determined by the ratio $\frac{\sigma_R}{\sigma_L}$, and does not depend on $N$. This results in a characteristic $R \propto N^{1/3}$ scaling law~\cite{Walker1990, Camara2014}, providing a clear signature of this regime defined by a spatially linear trapping force and weak attenuation of the trapping beams.

This simplified treatment of~\cite{Walker1990} was extended in~\cite{Labeyrie2006} in order to account for larger optical densities and to analyze MOT instabilities that can occur for larger atom numbers. This approach allowed to derive a criterion for the threshold of MOT instabilities that were observed when $N$ exceeded a certain critical value. In this model, which assumed a constant density, the nonlinear dependence of the attenuated trapping force on both position and velocity was retained. An unstable regime was found to occur when the cloud's radius was larger than a critical value $R_c$ given by:
\begin{equation} 
\mu~\nabla B~R_{c} \approx \left|\delta\right|
\label{threshold}
\end{equation}
where $\mu = 2\pi \times 1.4 \times 10^6$ s$^{-1}$G$^{-1}$ for the considered Rubidium transition. The MOT thus becomes unstable when the Zeeman detuning at the cloud's edge exceeds the absolute value of the laser detuning. In this situation, the total force at the cloud's edge reverses its sign, and the atomic motion becomes driven instead of damped. For $\delta = -2 \Gamma$ and $\nabla$B = 10 G/cm, the criterion given by Eq.1 yields $R_{c} \approx 9$ mm. Such a large MOT size typically requires a large atom number $N > 10^{10}$. While this simplified model provided an intuitive picture for the emergence of the instability and the existence of an instability threshold, it did not make quantitative predictions and was not able to describe the dynamics of the unstable cloud.

The assumption of a constant density was relaxed in~\cite{Pohl2006, Gattobigio2010}, based on a kinetic theory that described the phase space density of the atoms using a spatial radial symmetry hypothesis. The numerical test-particle simulations of the derived kinetic equations were able to confirm the simple instability criterion given by Eq.~\ref{threshold}. They also yielded a generic shape for the atomic density distribution of stable clouds under the form of truncated Gaussians. Finally, these simulations provided insights on the mechanism of the instability, and gave access to the dynamics of the cloud in the unstable regime. However, a limitation of this approach was the assumed spherical symmetry, effectively reducing the dimension of the problem to 1D and preventing e.g. center-of-mass motion in the dynamics of the unstable cloud.

\subsection{III.b. Three-dimensional numerical simulations}
To overcome the limitations of the previous models, we have developed a 3D numerical approach based on a microscopic description of the light-atom interaction. The detailed description of this model will be published elsewhere, we simply outline here its main features.

Since modeling $10^{11}$ atoms is out of reach with present computers, we used $N_s = 7 \times 10^3$ ''super-particles'', each representing $\alpha = \frac{N}{N_s}$ real atoms. The mass and scattering cross-section of these super-particles are $\alpha$ times larger than that of individual atoms. We checked the validity of this approach by verifying that the outcome of the simulation becomes independent of $N_s$ for $N_s > 5 \times 10^3$. What we simulate is the dynamics of these $N_s$ particles submitted to the three MOT forces mentioned earlier: the trapping force, the compressive attenuation force and the repulsive re-scattering force. The finite temperature of the cloud is accounted for by including a velocity diffusion term in the dynamics, which depends on the photon scattering rate for each particle. We use a leap-frog algorithm~\cite{Kaya1998} to compute the particles dynamics.

In the following, we describe the essential steps employed in the simulations. We use a Doppler model for the various forces, which are based on radiation pressure. To simplify the expressions given below, we assume here that our particles are two-level atoms. However, in the simulation we use a more realistic 0 $\rightarrow$ 1 transition model. It is obviously much simpler than the actual 2 $\rightarrow$ 3 transition of the D2 line of $^{87}$Rb used in the experiment, but allows for a correct description of the three-dimensional trapping by the MOT~\cite{Brickman2007}.

To calculate the forces acting on a particle located at position $r$, we first compute the local intensity $I(r)$ of each of the six laser beams. This is achieved by calculating their attenuation due to the rest of the cloud. By summing independently the radiation pressure forces $F_{rp}$ due to each beam, we obtain the trapping plus attenuation force. We thus neglect effects of cross-saturations between laser beams~\cite{Romain2011}, which is strictly valid only in the low saturation limit.

The radiation pressure exerted by one laser beam of intensity $I(r)$ on an atom located at position $\textbf{r}$ is of the form:
\begin{equation} 
F_{rp}(r,v)=\frac{\sigma_L(r,v) I(r)}{c}
\label{Frp}
\end{equation}
where $c$ is the speed of light. The scattering cross-section for a laser photon $\sigma_L$ is: 
\begin{equation} 
\sigma_L(r,v)=\frac{3\lambda^2}{2\pi}\frac{1}{1+I_{tot}(r)/I_s+(2 \delta_{eff}(r,v)/\Gamma)^2}
\label{sigma_L}
\end{equation}
where $\lambda$ is the wavelength of the atomic transition. The presence of the other laser beams is taken into account by the $I_{tot}(r)$ term in the denominator of expression~\ref{sigma_L}, which is the total local laser intensity obtained by summing the intensities $I_{i}(r)$ of each laser beam.  

Considering for simplification a beam propagating in the positive $x$ direction, the Doppler- and Zeeman-shifted detuning $\delta_{eff}$ is:
\begin{equation} 
\delta_{eff}(x,v)=\delta - \textbf{k}.\textbf{v} - \mu \nabla B x
\label{delta eff}
\end{equation}

The re-scattering force acting on the particle at position $r$ is obtained by summing all binary interactions with the other particles in the cloud. For a second particle located at position $r^{\prime}$, the binary interaction $F_{bin}(r,r^{\prime})$ is of the form:
\begin{equation} 
F_{bin}(r,r^{\prime})=\frac{I_{tot}(r^{\prime}) \sigma_L(r^{\prime}) \sigma_R(r,r^{\prime})}{4\pi c (r^{\prime} - r)^2}
\label{binary}
\end{equation}
The computation of the re-absorption cross-section $\sigma_R$ is not straightforward. It involves the convolution of two quantities: the absorption cross-section for a scattered photon with a certain frequency by the atom illuminated by the total laser field of intensity $I_{tot}(r)$, and the spectrum of the light scattered by the atom at position $r^{\prime}$~\cite{Walker1990, Pruvost2000}. Indeed, the atom is a nonlinear scatterer for light, and the scattering process is inelastic if the saturation parameter is not negligible compared to unity. We compute both the scattered light spectrum and the absorption cross-section for scattered light using the approach developed by Mollow~\cite{Mollow1969,Mollow1972}. Note that since we take into account the inhomogeneous laser intensity distribution inside the cloud, $\sigma_R$ depends on both spatial coordinates $r$ and $r^{\prime}$, and thus plays the role of a nonlocal effective charge in the Coulombian binary interaction described by Eq.~\ref{binary}.

\begin{figure}
\begin{center}
\includegraphics[width=1\columnwidth]{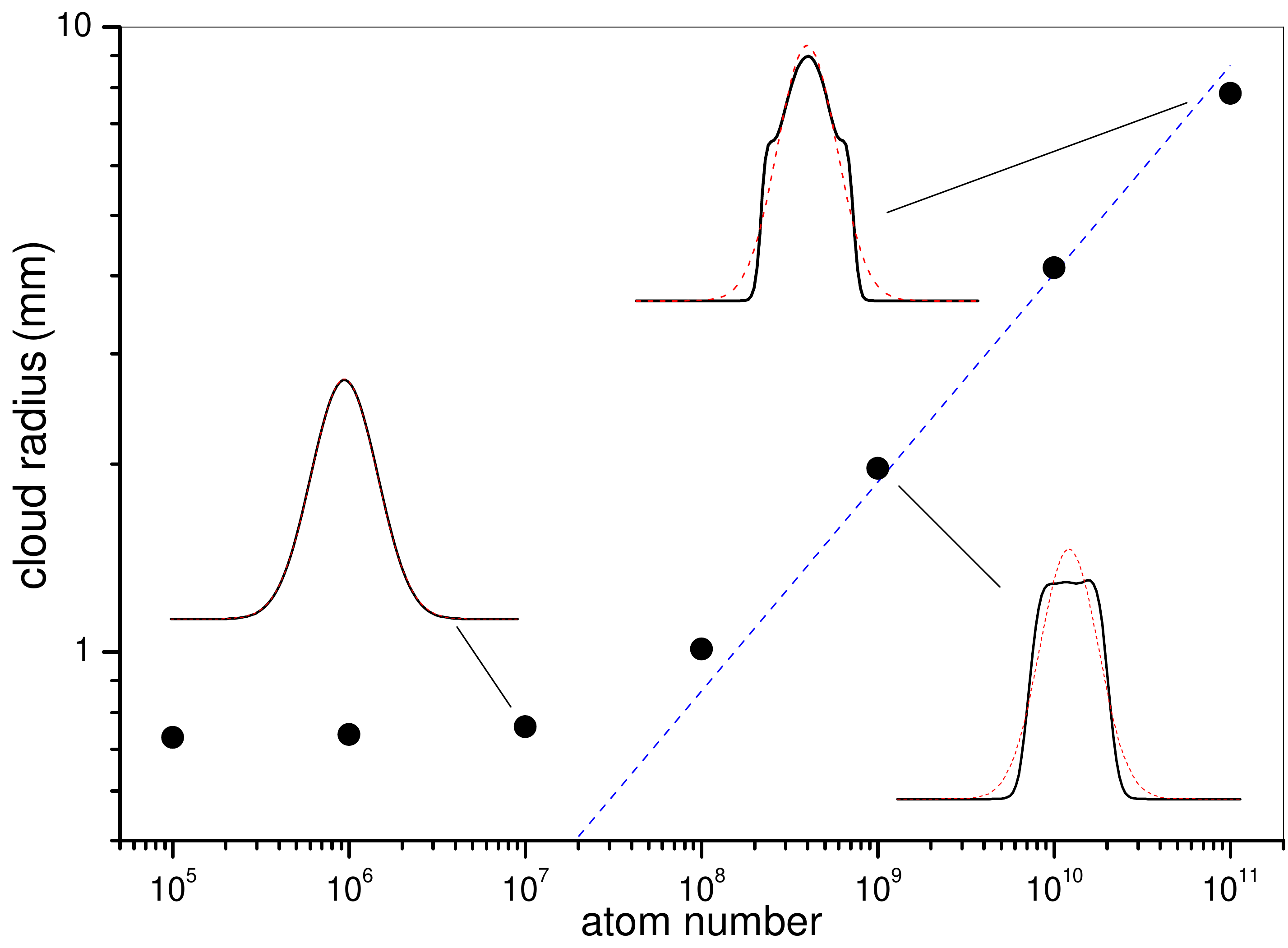}
\caption{Simulated MOT size versus $N$. The parameters are: $\delta = -5.5\Gamma$, $\nabla$B = 7.2 G/cm and I = 5 mW/cm$^2$. We plot the cloud's \textit{rms} radius versus $N$ (dots). The dashed line corresponds to the $N^{1/3}$ scaling. The insets show examples of cloud density profiles which have been spatially rescaled. The solid curves are the density profiles while the dashed curves correspond to Gaussian fits.}
\label{sizesSim}
\end{center}
\end{figure}

An illustration of the result of this simulation is provided in Fig.~\ref{sizesSim}, where we plot the cloud's \textit{rms} radius versus the number of simulated atoms, in the stable regime. As can be observed, below $10^7$ atoms the cloud size is $N$-independent and its profile is Gaussian as expected in the temperature-limited regime. For $N > 10^8$, the cloud's radius increases as $N^{1/3}$ (dashed line), the scaling predicted by Ref.~\cite{Walker1990}. The increase of cloud size with $N$ is a clear signature of the multiple-scattering regime. Within this regime, we observe variations of the cloud's density profile. Around $10^9$ atoms, the profile displays a rather flat top. At even higher atom numbers, the top of the profiles rounds off and gets closer to a Gaussian. In all instances, however, the wings of the profiles in the multiple-scattering regime are decaying faster than a Gaussian. Overall, the observed evolution is consistent with our previous observations~\cite{Camara2014}, and also with the theoretical predictions of Ref.~\cite{Pohl2006}. It should be noted that the simulated clouds are systematically larger than those observed in the experiment, roughly by a factor 2. We discuss the implications of this observation in section V.

\begin{figure}
\begin{center}
\includegraphics[width=1\columnwidth]{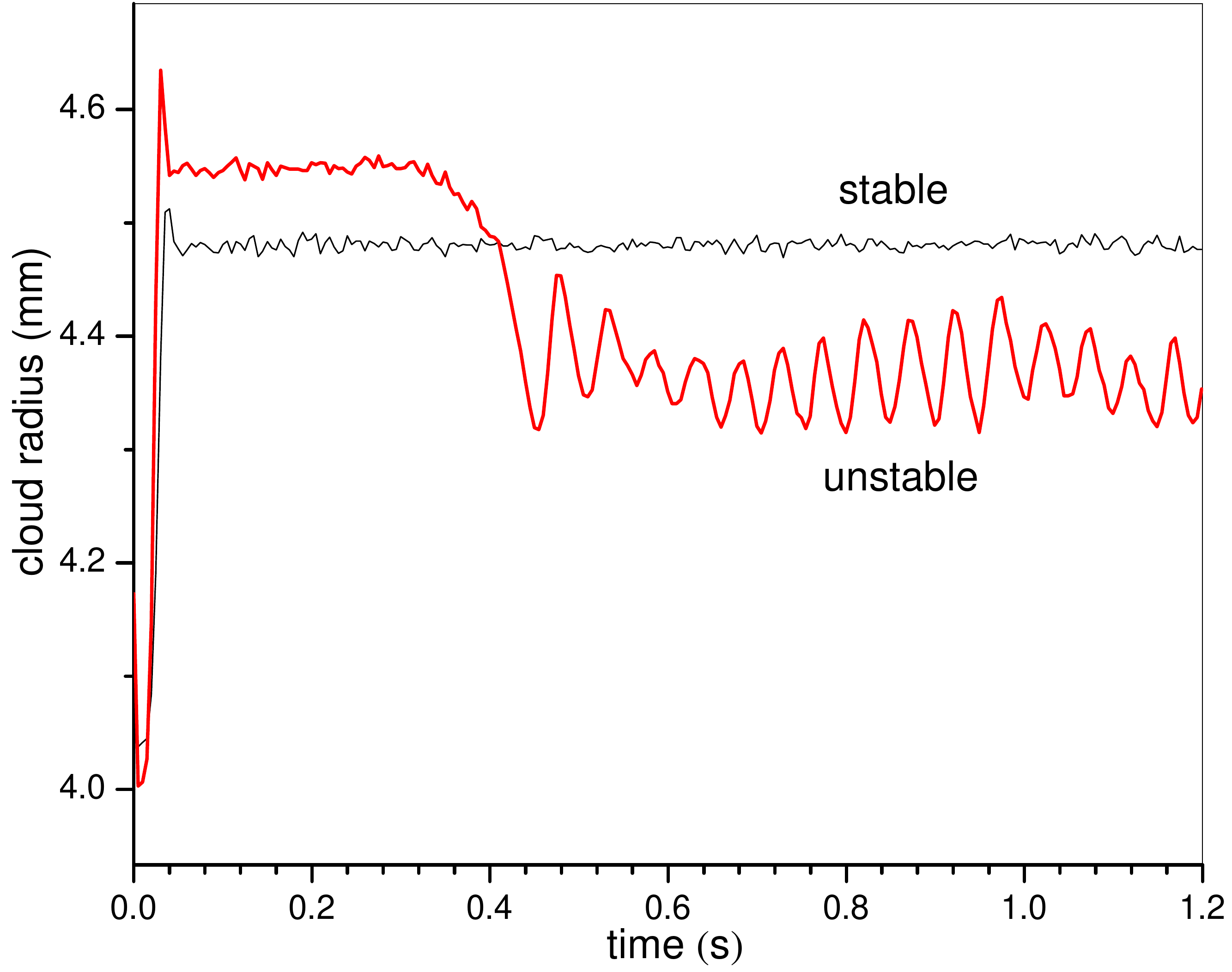}
\caption{Instability of simulated MOT. We plot the cloud's \textit{rms} radius versus time, in the stable regime ($\delta = -3.0 \Gamma$, thin curve) and in the unstable regime ($\delta = -2.8 \Gamma$, bold curve). The parameters are: $N = 1.5 \times 10^{10}$, $\nabla$B = 3 G/cm, $I = 5$ mW/cm$^2$.}
\label{InstabSim}
\end{center}
\end{figure}

The numerical simulations not only reproduce the behavior of a stable MOT in the multiple scattering regime, but also and most importantly they yield an unstable behavior for parameters close to those used in the experiments. The onset of the instability as observed in the simulations is illustrated in Fig.~\ref{InstabSim}. We observe a sharp transition between stable and unstable behaviors when the control parameter, here is the laser detuning, is varied. We plot the temporal evolution of the \textit{rms} radius of the cloud both below ($\delta = -3\Gamma$) and above ($\delta = -2.8\Gamma$) threshold. The initial transient (first 50 ms) is due to the slight mismatch in size and shape between the Gaussian atomic distribution used as a starting point for the simulation, and the final distribution. In the unstable regime, we generally observe another transient before the onset of oscillations, whose duration depends on the distance from threshold. In Fig.~\ref{InstabSim}, this duration is roughly 0.4 s.

\section{IV. Experimental results and comparison with simulations}
\label{results}

\subsection{IV.a. Experimental determination of instability threshold}

\begin{figure}
\begin{center}
\includegraphics[width=1\columnwidth]{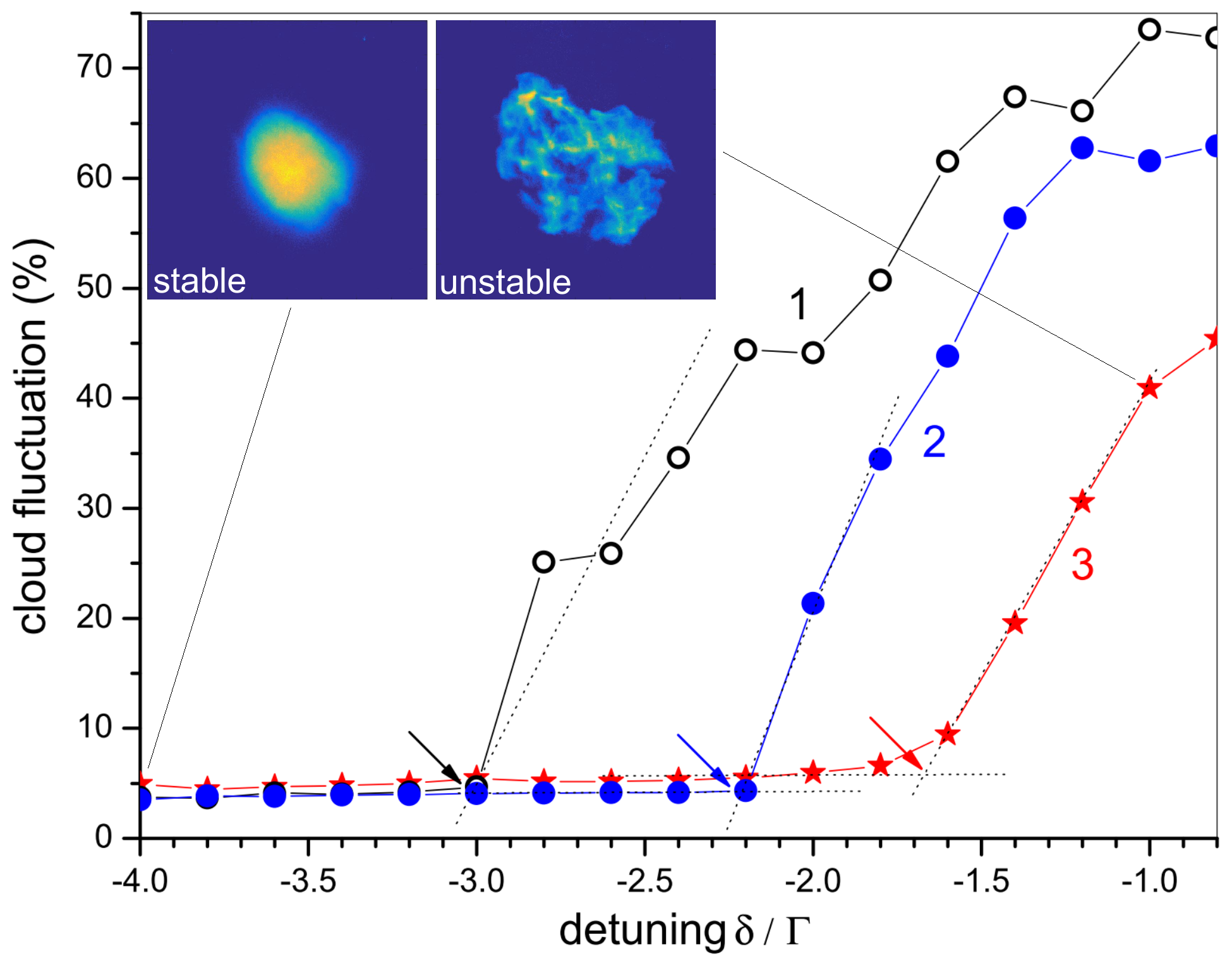}
\caption{Determination of instability threshold. We plot the cloud fluctuation (see text) versus detuning for three values of magnetic field gradient: 1) $\nabla$B = 12 G/cm (open circles), 2) $\nabla$B = 4.8 G/cm (dots), and 3) $\nabla$B = 1.2 G/cm (stars). An extrapolation of the observed growth rate (dotted lines) allows to determine the threshold detuning (arrows). The insets show examples of fluorescence images for a stable and an unstable cloud.}
\label{ThreshDet}
\end{center}
\end{figure}

To determine the instability threshold, we monitor the evolution of the cloud's spatial density distribution during the dynamics. As discussed in Section II, only one image is recorded during each experimental cycle described in Fig.~\ref{setup}B, which corresponds to a random probing of the dynamics of the cloud. We thus record a given set of typically 50 images, and then compare the images two by two. This is done by subtracting the two images, and spatially integrating the squared difference image. After normalization, this operation yields the ''cloud fluctuation'' of Fig.~\ref{ThreshDet}, a number whose value is zero if the two images are identical, and one if there is no overlap between the two density distributions (corresponding to a maximal deformation). The operation is repeated for many pairs of images and the corresponding fluctuations are averaged.

Fig.~\ref{ThreshDet} illustrates the behavior of the cloud fluctuation as $\delta$ is varied over the whole experimental range $-4\Gamma \leq \delta \leq -0.8\Gamma$, for the three values of $\nabla$B. As can be seen, crossing the threshold results in an abrupt increase of the fluctuation. The position of the threshold can be estimated by fitting the initial growth by a linear function and extrapolating to the value below threshold.

\subsection{IV.b. Threshold detuning versus atom number}

In this section we study the impact of $N$ on the instability threshold. To this end, we vary the number of trapped atoms by adjusting the diameter of the MOT beams using diaphragms. This provides an efficient way of tuning $N$ without affecting the other MOT parameters, as demonstrated in~\cite{Camara2014}.

\begin{figure}
\begin{center}
\includegraphics[width=1\columnwidth]{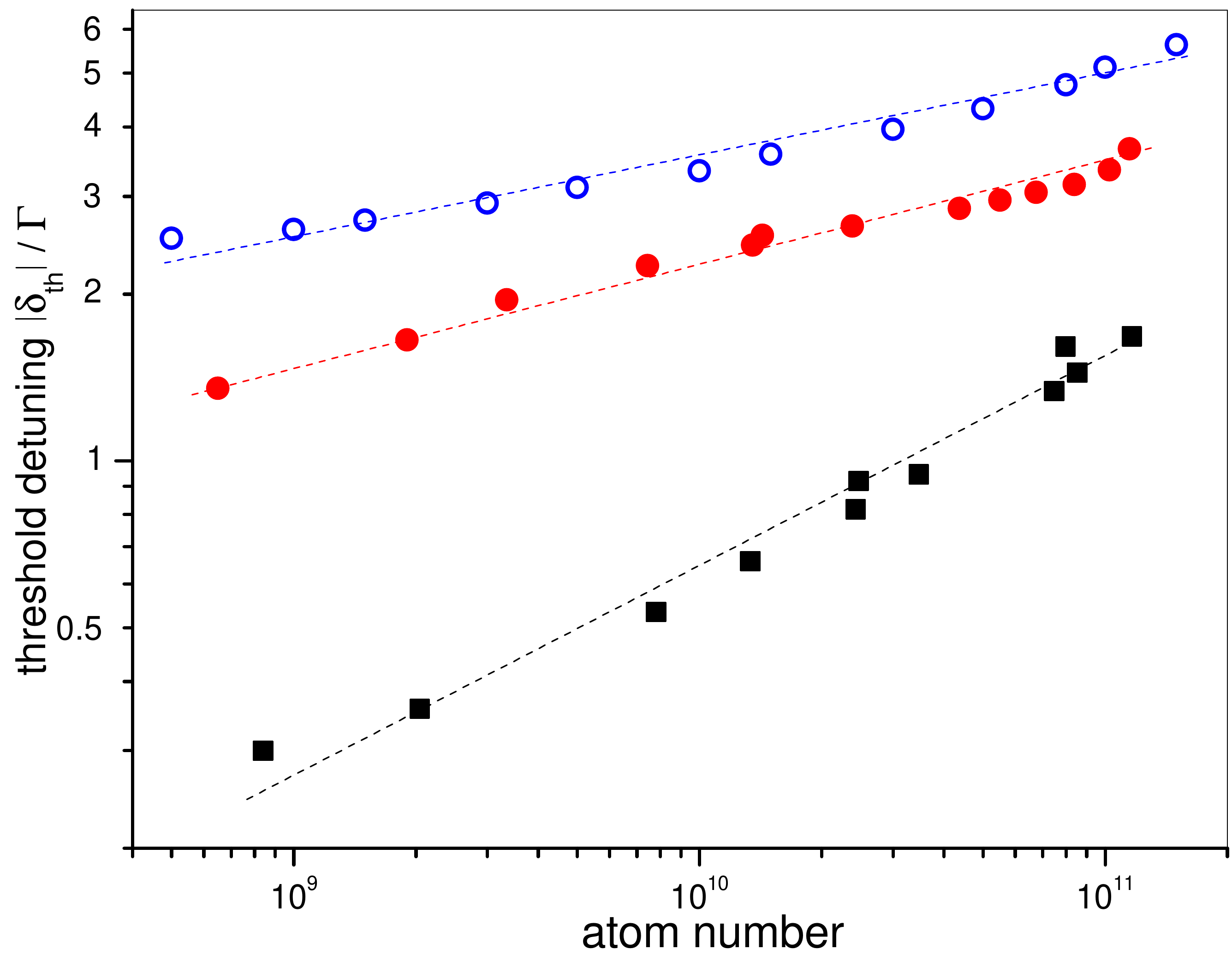}
\caption{Threshold detuning (absolute value) versus atom number. The dots are experimental data for a magnetic field gradient $\nabla$B = 7.2 G/cm and a beam intensity $I = 5$ mW/cm$^2$. The circles are the simulation result for the same parameters. The squares correspond to the prediction of the model of Ref.~\cite{Labeyrie2006} using experimentally measured cloud sizes (see text). The dotted lines correspond to linear fits of the log-log data.}
\label{deltavsN}
\end{center}
\end{figure}

The experimental variation of $\delta_{th}$ with $N$ is reported in Fig.~\ref{deltavsN} (dots), in log-log scale. Very roughly, we observe that $\left|\delta_{th}\right|$ increases by $\Gamma$ when $N$ increases by one order of magnitude. A linear fit of the data in the log-log plot yields a slope of 0.17 (dotted line). The result of the numerical simulations described in section III.b is reported in circles. It shows a similar scaling, but with an offset of approximately $\Gamma$.

We also show the prediction of the model of Ref.~\cite{Labeyrie2006} (squares), as given by Eq.~\ref{threshold}. This equation establishes a link between $\nabla B, \delta$ and the critical radius $R_c$, but does not provide an expression for the cloud size as a function of $N$. Thus, we use the cloud sizes measured in the experiment to compute $\delta_{th}$ using Eq.~\ref{threshold}. However, the model of~\cite{Labeyrie2006} assumes a constant density while the experimental profiles are closer to Gaussians. We thus have to choose a definition of $R_c$ to use in Eq.~\ref{threshold}. In Fig.~\ref{deltavsN}, we used $R_c = 2 \sigma$, where $\sigma$ is the measured \textit{rms} cloud size in the plane of the magnetic field gradient coils. Since we find a scaling $\sigma \propto N^{0.36}$, the prediction Eq.~\ref{threshold} is definitely different from both the experimental observation and the result of the numerical simulations. This is a clear indication that the model used in~\cite{Labeyrie2006} cannot quantitatively describe the dependence of the instability threshold on the atom number.

\subsection{IV.c. Threshold detuning versus magnetic field gradient at fixed atom number}

\begin{figure}
\begin{center}
\includegraphics[width=1\columnwidth]{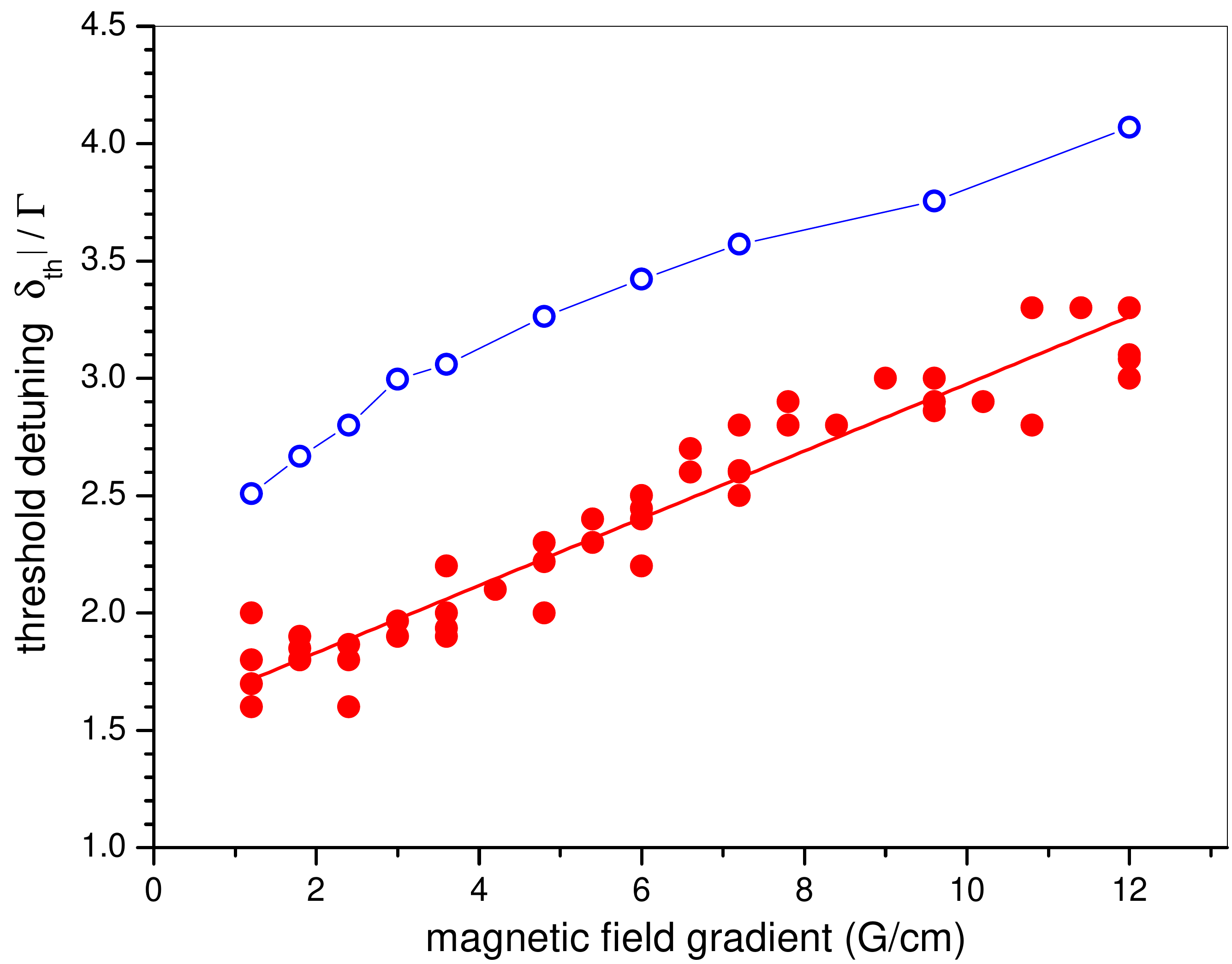}
\caption{Threshold detuning versus magnetic field gradient at fixed atom number. The experimental data (including several runs) is shown as dots, with a linear fit (solid line). The parameters are: N = $1.5 \times 10^{10}$, $I = 5$ mW/cm$^2$. The circles correspond to the numerical simulation result.}
\label{compNfixed}
\end{center}
\end{figure}

We now present measurements of the threshold detuning $\delta_{th}$ as $\nabla$B is varied, using the procedure described in section II to maintain a fixed number of trapped atoms N = 1.5 $\times~10^{10}$. The results of different experimental runs are shown as dots in Fig.~\ref{compNfixed}, together with a linear fit taking into account all the data (solid line). The threshold detuning is seen to increase (in absolute value) linearly with $\nabla$B, with a slope $\approx~0.14~\Gamma/$(G/cm).

We compare the experimental data to the result of the numerical simulations (circles) using the experimental parameters. Once again, we observe that the slopes of both experimental and simulated curves are very similar. This agreement is a good indication that our numerical model captures efficiently the main ingredient involved in the determination of the instability threshold. We note again that the simulated thresholds are systematically larger (in absolute value) than the experimental ones by approximately $\Gamma$. 

\section{V. Discussion}
\label{discussion}
The new experimental data reported in this work need to be compared to our earlier results of~\cite{Labeyrie2006}. They are obtained in conditions that are both better controlled (threshold at constant $N$) and more extended (threshold as a function of $N$), and thus put more stringent constraints on the models they are being compared to. In particular, the simple analytical model of~\cite{Labeyrie2006}, which seemed in reasonable agreement with the early experimental data, is now clearly unable to reproduce the scaling observed when the atom number is varied. Although we have no reason to question the physical picture of the instability mechanism that it conveys, it is too simplified to reproduce even qualitatively the experimentally observed behavior.    

This is not the case for the numerical simulations, which are in good qualitative agreement with both experiments at fixed $N$ and as a function of $N$. This indicates that our improved model catches the important ingredients determining the behavior of the MOT instability threshold. We plan in the future to investigate which of these ingredients are determinant in describing the correct MOT behavior.

The constant offset of about one $\Gamma$ in the threshold detuning between experiment and numerics is probably linked to the larger cloud sizes found in the simulations. The origin of this mismatch is not at present clearly identified, but it is not surprising considering the large number of simplifications still included in the model. The most prominent is the simplified atomic structure, possibly yielding a different effective Zeeman shift in the MOT.

\section{VI. Conclusion}
We presented in this paper a detailed experimental and numerical study of the instability threshold for a balanced, six-independent beams magneto-optical trap containing large numbers of cold atoms. Using an improved experimental scheme, we were able to study the impact of the atom number on the threshold. We also measured the ($\delta$, $\nabla$B) unstable boundary while maintaining this atom number fixed. These experimental results were compared to a three-dimensional numerical simulation of the MOT based on a microscopic description. We obtain a good qualitative agreement, despite some unavoidable simplifications in the description of the MOT physics. The scaling of the threshold with atom number, for both experiment and simulations, is clearly different from that given by the analytical approach of~\cite{Labeyrie2006}. Our numerical model also allows us to go beyond the approach of~\cite{Pohl2006}, which was assuming a central symmetry. In particular we now can and do observe center-of-mass as well as radial oscillations. This approach will be useful in the future to investigate the atom cloud dynamics in the unstable regime.

\section{Acknowledgements}
Part of this work was performed in the framework of the European Training Network ColOpt, which is funded by the European Union (EU) Horizon 2020 programme under the Marie Sklodowska-Curie action, grant agreement No. 721465.

\end{document}